\documentclass[conference]{IEEEtran}
\IEEEoverridecommandlockouts
\usepackage{cite}
\usepackage{braket}
\usepackage{amsmath,amssymb,amsfonts} 
\usepackage{algorithm}
\usepackage[utf8]{inputenc}
\usepackage{array}
\usepackage{colortbl}
\usepackage{tabularray}
\usepackage{rotating}
\usepackage{color}
\usepackage{svg}

\usepackage{graphicx}
\usepackage{textcomp}
\usepackage{xcolor}
\usepackage{gensymb}
\usepackage{subfigure}
\usepackage{epsfig}
\usepackage{epstopdf}
\usepackage{array}
\usepackage{makecell} 
\usepackage{makecell}
\usepackage{textcomp}
\usepackage{booktabs}
\usepackage{threeparttable}
\usepackage{booktabs}
\usepackage{caption}
\usepackage{multirow}
\usepackage{algorithm}
\usepackage{algpseudocode}
\usepackage{amsmath}

\def\BibTeX{{\rm B\kern-.05em{\sc i\kern-.025em b}\kern-.08em
    T\kern-.1667em\lower.7ex\hbox{E}\kern-.125emX}}
\begin{document}

\title{Towards 6G Digital Twin Channel Using Radio Environment Knowledge Pool
 \\
\thanks{Jialin Wang and Jianhua Zhang are co-first authors
(Corresponding author: Yuxiang Zhang).}
}

\author{\IEEEauthorblockN{Jialin Wang, Jianhua Zhang, Yuxiang Zhang, Yutong Sun, Gaofeng, Nie, Lianzheng Shi, Ping Zhang, Guangyi Liu}
\IEEEauthorblockA{
}
}

\maketitle

\begin{abstract}
The digital twin channel (DTC) is crucial for 6G wireless autonomous networks as it replicates the wireless channel fading states in 6G air interface transmissions. 
It is well known that the physical environment influences channels. 
A key task for accurately twinning channels in complex 6G scenarios is establishing precise relationships between the environment and the channels.
In this article, the radio environment knowledge pool (REKP) is proposed, with its core function being to construct and store as much knowledge between the environment and channels as possible.
Firstly, the research progress related to DTC is summarized, and a comparative analysis of these achievements on key indicators in digital twin is conducted, proposing the challenges faced in knowledge construction.
Secondly, instructions on how to construct and update REKP are given.
Then, a typical case is presented to demonstrate the great potential of REKP in enabling DTC.
Finally, how to utilize REKP to address open issues in the 6G wireless communication system is discussed, including enhancing performance, reducing costs, and keeping a trustworthy DTC.

\end{abstract}


\section{Introduction}

The 6G goals and framework defined by the Working Party 5D (WP 5D) of ITU Radiocommunication Sector (ITU-R), determining six major scenarios and fifteen performance indicators  \cite{ITUFramework2023}. 
To achieve these goals, 6G will construct a new communication network paradigm with 
space–aerial–terrestrial–ocean integration, and bring emerging technologies, such as ultra-massive multiple-input multiple-output, integrated sensing and communication (ISAC), and new spectrum (terahertz, optical), to fulfill the sophisticated requirements.
Therefore, 6G faces a more complex communication environment and channel changes than 5G, which brings greater challenges to 6G channel research.

The channel is the medium between the transmitter (TX) and the receiver (RX). 
Its properties determine the ultimate performance limit of a wireless communications system (WCS) \cite{ZhangChannel2023}.
Therefore, channel modeling is a pivotal aspect explored in every generation of wireless communication to support system deployment and evaluation. 
Currently, statistical channel modeling is the mainstream method.
Firstly, channel measurements are conducted considering typical scenarios such as urban microcell (UMi), urban macrocell (UMa), rural macrocell (RMa), and indoor hotspots (InH).
Then, statistical channel modeling and simulation are derived based on these measurement results.
However, the generality of offline channel models derived through constructing these typical scenarios is constrained.
Furthermore, 6G faces greater bandwidth, larger antenna scales, and more scenarios, frequency bands, and technologies.
WCS necessitates channel estimation and feedback in such a communication environment, resulting in significant system overhead.
Therefore, it is a challenging task to obtain a high-precision and low-overhead channel for 6G WCS.
To pursue optimal communication system performance, a novel approach to channel rapid acquisition is considered under the development of artificial intelligence and converged sensing and communication techniques.

\begin{figure*}[!t]
\centerline{\includegraphics[scale=0.4]{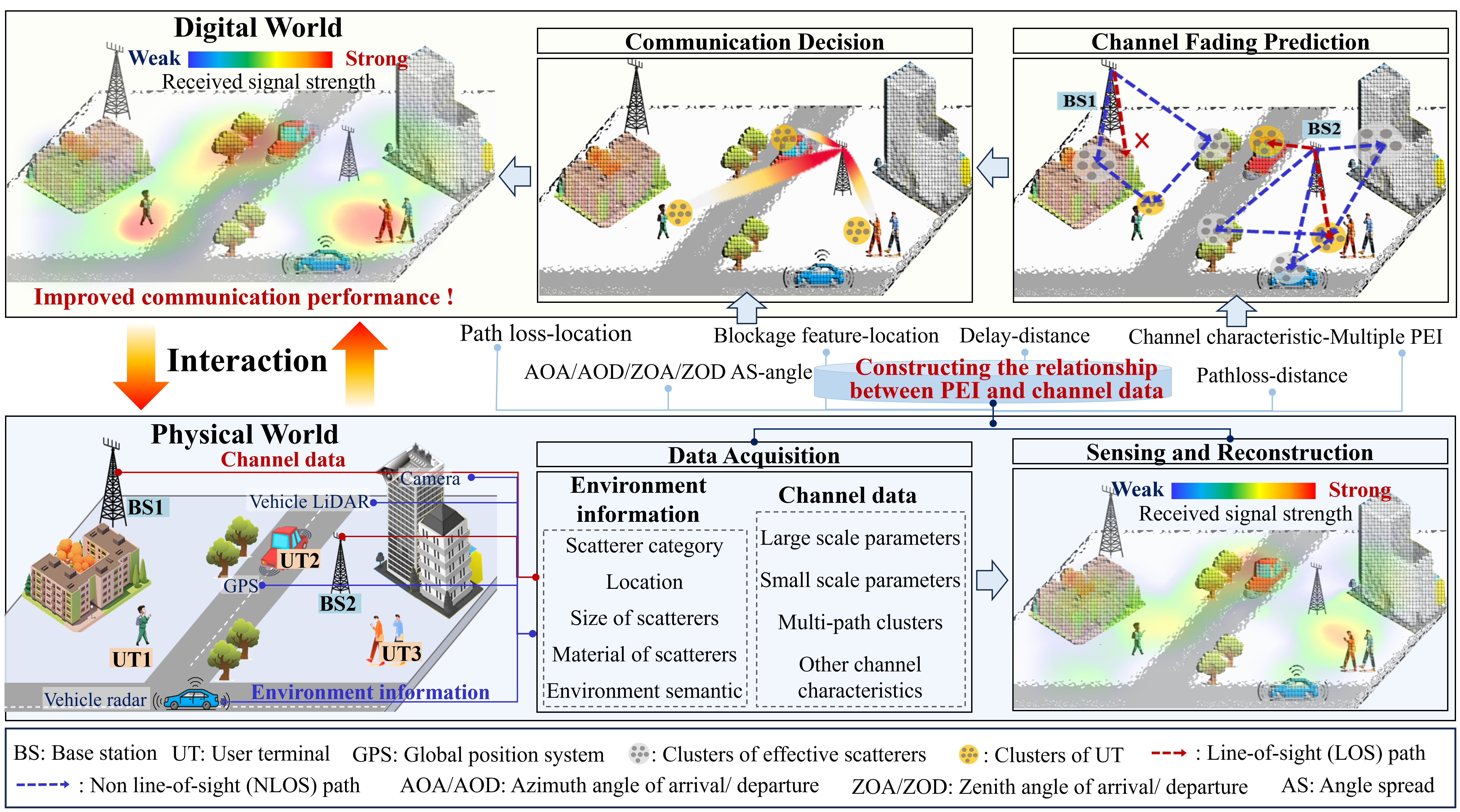}}
\caption{DTC implementation framework in that wireless channels in the PW map into the DW, including sensing and reconstruction, channel fading prediction, and communication decision by constructing the relationship.}
\label{Fig1}
\end{figure*}

Digital twin (DT) is one of the enabling technologies for the 6G communication system \cite{AlkhateebReal2023}.
Considering the controllable, replicable advantages of DT \cite{Multi-Spectrum2023, GaoDigital2023}, and further emphasizing the interaction between the channel and the WCS, we define the digital twin channel (DTC) for the first time.
DTC is a digital virtual mapping of a wireless channel that reflects the entire process of channel fading states and variations in the physical world (PW), which continually interacts with the WCS channel requirements to support WCS in actively adapting to complex communication environments.
Based on the capabilities of DTC and the predictive 6G network proposed by our team in previous work \cite{NiePredictive2022}, a novel DTC implementation framework is feasible, as shown in Fig. \ref{Fig1}.
Firstly, propagation environment information (PEI) and channel data are captured from the PW for environment sensing and reconstruction. 
Then, channel fading prediction and communication decisions are performed in the digital world (DW) by constructing the relationship between PEI and channel data. 
The decision result is fed back from the DW to the PW, and the real-time PEI in the PW is combined to realize the interaction.

It is well known that the propagation environment influences the channel. 
The radio wave is emitted from the TX and interacts with various scatterers in the propagation path through reflection, scattering, and diffraction, then reaches the RX. 
The radio waves at RXs show the characteristics of multipath clustering. 
The scatterers' shape, position, and material affect the radio wave propagation characteristics.
Therefore, some studies focus on the relationships in wireless channel changes affected by PEI \cite{ZhangCN2016, MiaoDemo2023, ZengToward2021, SunHow2023}.
An in-depth representation of the relationship between the PEI and the channel can improve the interpretability of the DTC implementation framework, which is the critical metric of DT \cite{ZhuIntelligent2023}.
Therefore, the mapping relationship between the wireless environment and channel must be solved to realize DTC, especially understanding the essence, relationship, and rules between PEI and channel data to improve the interpretability and build a universal channel generation process.

In this article, a radio environment knowledge pool (REKP) is proposed to serve as a specific enabler for DTC.
radio environment knowledge (REK) is the representation of the essence, relationships, and regular patterns between radio propagation and its environment, derived through the utilization of PEI and channel data obtained in the PW.
The core function of REKP is to construct and store as much REK between PEI and channel data as possible rather than being tailored to one specific type of scenario.
It plays a crucial role in wireless channels in the PW mapping into the DW with the advantages of being controllable, interpretable, renewable, and generalized.
REKP enhances WCS's adaptability to the communication environment and is crucial in achieving DTC.

The organization of this article is as follows.  
The related works about the DTC are summarized in Section \ref{secII}. 
Section \ref{secIII} discusses various building modules for building REKP.
A path loss knowledge construction and the case based on this knowledge are given in Section \ref{secV}.
Open issues for DTC achievement are discussed in Section \ref{secVi}.
Section \ref{secVI} gives the conclusion and future work.

\section{Related work}\label{secII}

\begin{table*}[htbp]
\setlength{\tabcolsep}{11pt} 
\renewcommand\arraystretch{1.2}  
\caption{Comparative analysis of research progresses about efficiency, cost, data security, and interpretability.}
\label{related_work_summary_table}
\begin{center}
\begin{tabular}{|m{0.25cm}<{\centering}|m{2.5cm}<{\centering}|m{3.2cm}<{\centering}|m{1.2cm}<{\centering}|m{2cm}<{\centering}|m{1.8cm}<{\centering}|m{0.8 cm}<{\centering}|}
\hline
 {\textbf{Ref.}} & {\textbf{Application}} & {\textbf{Efficiency}}& {\textbf{Cost}} & {\textbf{Data security}} & {\textbf{Interpretability}} &  {\textbf{Mode}}\\
\hline
 {\cite{AlkhateebReal2023}} & {Real-time DT vision} & {Reduce system overhead} & {EC \& MT} & {Measurement \& UT devices} & {EPT \& ML}  & {Online} \\ 
\hline
    {\cite{Multi-Spectrum2023}} & {Channel modeling}  & {Reduce pathfinding time} & {EC \& HD} & {Simulation} & { EPT } &  {Offline} \\
\hline
 {\cite{GaoDigital2023}} & {Radio testing twin} & {Reduce test run time} & {EC \& HD} & {Simulation \& UT devices} & {EPT} & {Online} \\
\hline
 {\cite{ZhangCN2016}} & {Channel modeling} & {Improve channel accuracy} & {DA} & {Simulation \& measurement} & {EPT \& ML} &  {Offline} \\
\hline
 {\cite{MiaoDemo2023}} & {Scenario-oriented DTC} & {Improve environment adaptability} & {EC \& HD} & {Simulation \& UT devices} & {EPT \& ML} & {Online}  \\
\hline
 {\cite{ZengToward2021}} & {Channel map} & {Training-free} & {EC \& DA} & {Simulation \& UT devices} & {ML} &{Offline} \\
\hline
 {\cite{SunHow2023}} & {Channel prediction} & {Reduce training time} & {MT} & {Simulation} & {ML} &  {Offline} \\
\hline
\multicolumn{7}{|l|}{DA: Data acquisition; EC: Electromagnetic calculation; EP: Electromagnetic propagation.  }\\
\multicolumn{7}{|l|}{HD: Hardware devices; ML: Machine learning; MT: Model training.}\\
\hline
\end{tabular}
\end{center}
\end{table*}

\subsection{Research status related to DTC}

Researchers extensively study DT for the channel, which is crucial in the physical layer and channel research. 
DT for channel alleviates the burden of channel acquisition and reduces guide frequency overhead, thereby optimizing WCS efficiency \cite{AlkhateebReal2023}.
Wireless radio propagation paths from microwave to visible light are reconstructed utilizing DT technology. 
The results show that the MAEs are around 2 to 10 dB less than the baseline method, around 3000 times faster than the ray tracing-only method \cite{Multi-Spectrum2023}.
Regarding channel testing, utilizing DT enables wireless radio testing to be conducted in a repeatable DW, simulating real-world operations \cite{GaoDigital2023}.

The development of artificial intelligence, Internet of Everything communication, and sensing techniques accelerates the implementation of DTC. 
Researchers explore the relationship between the environment and the channel to facilitate channel reconstruction in support of DTC.
A novel cluster-nuclei (CN)--based channel model is proposed to analyze the interplay between clusters and scatterers in the propagation environment, enabling channel reconstruction across various scenarios \cite{ZhangCN2016}.
Furthermore, leveraging computer vision and radar techniques, a channel prediction platform enhanced by sensing capabilities is developed to facilitate dynamic channel reconstruction in indoor settings \cite{MiaoDemo2023}.
A task-oriented site-specific database is curated through environment sensing, establishing a linkage between environment location and channel \cite{ZengToward2021}.
A direct mapping approach is employed to correlate channel data with environment attributes, thereby constructing environment semantics for channel prediction with getting 0.92 precision and saving over 87\% time cost \cite{SunHow2023}.

\subsection{Summary and challenges}

The research progress covers physical layer and channel studies based on the DT techniques or aims at achieving DTC. 
Table \ref{related_work_summary_table} provides a comparative analysis of these achievements' efficiency, cost, data security, and interpretability, which are key indicators of DT.
However, current research on interpretability remains limited to the relationship between environment and channel data, mainly using intelligent algorithms for exploration.
Understanding the essence of PEI and channel data can enhance the interpretability of DTC, but it still faces the following three challenges:
\begin{itemize}
  \item The knowledge is various pattern representations derived from data-rich PEI and channel data. Legitimate acquisition of high-quality data is a challenge.
  \item Describing how REK is constructed and the mutual influence between different pieces of knowledge for the interpretability of DTC is a challenge.
  \item A good solution involves knowledge sorting, updating, and removing to ensure the accuracy and relevance of the knowledge pool is a challenge.
\end{itemize}

\section{Construction of Radio Environment Knowledge Pool} \label{secIII}
The primary task of REKP is to characterize and store the feature relationships that exist in various forms between the environment and channel to enhance interpretability and generalization in achieving DTC. Compared to existing methods, the proposed REKP has advancements:
\begin{itemize}
  \item \textbf{Controllable:} REKP evolves from uncovering features and mapping relationships between PEI and channel data to investigating regular patterns, enabling a shift from uncertain to controllable relationships.
  \item \textbf{Interpretable:} The solidified knowledge between the environment and the channel is interpreted digitally in REKP by theoretical derivation, structured analysis, environment semantic representation \cite{SunHow2023}, etc., forming a more precise mapping.
  \item \textbf{Renewable:} REKP updates based on feedback from PEI. Internal feedback and updates occur in the knowledge representation module. When a new channel feature arises, the knowledge is transferred from related channel features, enabling external feedback and updating.
  \item \textbf{Generalized:} Controllable relationships, interpretable mappings, and renewable mechanisms make REKP exhibit greater adaptability to new channel features and communication tasks, which enables a generalization REKP.
\end{itemize}

\begin{figure*}[!t]
\centerline{\includegraphics[scale=0.4]{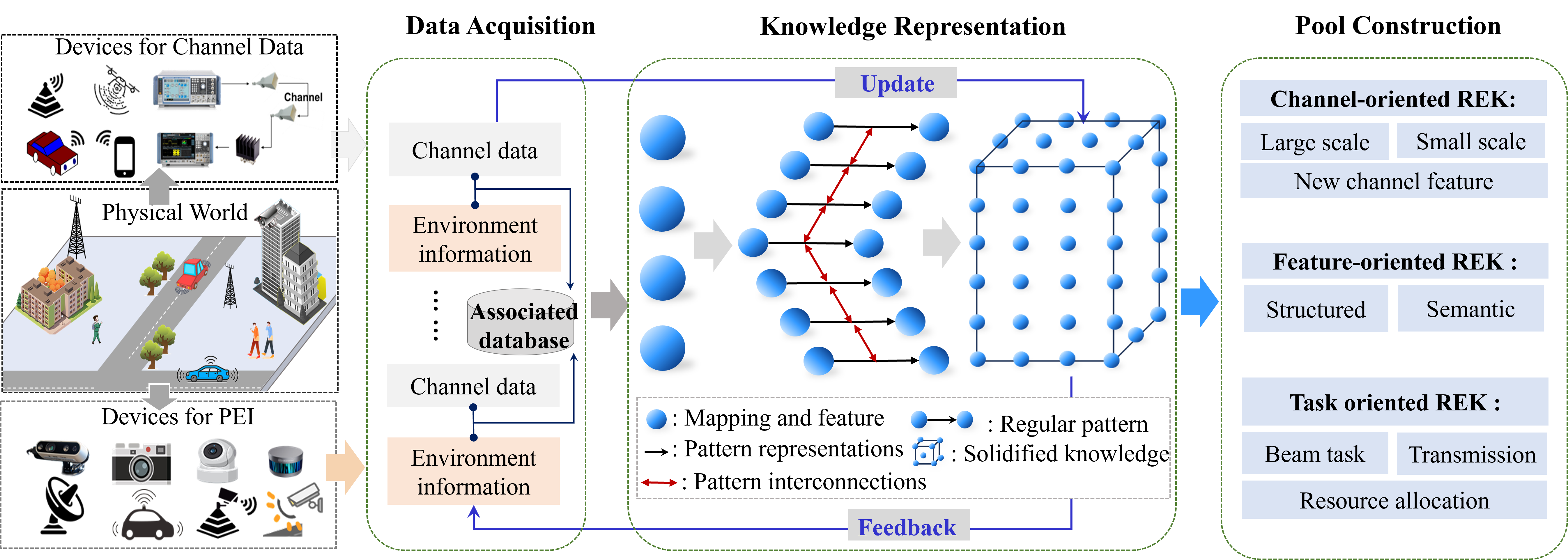}}
\caption{Architecture of radio environment knowledge pool.}
\label{Fig2}
\end{figure*}

The architecture of REKP consists of three modules, including data acquisition, knowledge representation, and pool construction, respectively, as shown in Fig. \ref{Fig2}.

\subsection{Data acquisition}
The data acquisition from the PW encompasses channel data and PEI. 
Some configuration details are collected from various infrastructures, including frequency band, bandwidth, antenna specifications, TX positions, and the precise positioning data of UTs such as mobile phones, vehicles, and machinery.
Authentic channel data is obtained through meticulous channel measurements.
These measurement results are processed to derive parameters such as channel impulse response, angle of arrival, delay spread, and received power.
Concurrently, the acquisition of multi-modal PEI is integral and facilitated by various sensing devices. For instance, RGB images are captured by 2D cameras, panoramic images by 360-degree cameras, depth information by depth cameras, and point cloud data by Lidar and 3D cameras.

A radio environment-associated database is constructed using these multi-modal PEI and channel data. 
Temporal and spatial alignment of PEI and channel data is achieved by integrating timestamps into the data sequences for temporal alignment and including position parameters of data collection devices for spatial alignment. 
The organized, labeled data generate correlated data pairs representing environmental conditions and channel characteristics at identical temporal points and location coordinates, all stored within the database.

\subsection{Knowledge representation}\label{secIIIB}

Knowledge representation is uncovering the inherent relationships between the environment and the channel.
Unlike previous studies focusing on the one-to-one or one-to-many feature mapping between PEI and channel parameters, knowledge representation focuses on the relationships and interactions among features. 

Firstly, the PEI features of the effective scatterers and channel features are excavated from fundamental features to intermediate recessive features and finally to advanced semantic features by relationship mapping or feature extraction. 
Effective scatterers refer to the scatterers that produce paths and influence the channel characteristics \cite{SunHow2023}. 
Fundamental features include geometric and statistical environmental features, offering a general understanding. Intermediate features involve edge and pixel features, abstracting information further. 
Advanced semantic features encapsulate channel and environment semantics.
Then, regular patterns between multi-domain, multi-scale channel features, and multi-modal PEI features are deduced, solidifying knowledge through pattern representations and interconnections.  
Unlike existing studies, the shift from feature mapping to knowledge representation emphasizes understanding feature relationships and their impacts.
Further, internal feedback and updates refine knowledge, which is detailed in Section III-D.

\subsection{Pool construction}
Currently, three categories of knowledge are constructed in REKP, which can be expanded upon or used as a foundation to derive new knowledge.


\textbf{Channel-oriented REK:} 
This category focuses on channel characteristics, including large-scale and small-scale parameters and multipath cluster characteristics.  
For instance, it addresses environment errors resulting from scatterer displacements, elucidating how environment uncertainty impacts the channel \cite{WangEffects2021}.  
Novel channel characteristics like shared knowledge in ISAC channels \cite{ZhangIntegrated2023} and electromagnetic response knowledge in re-configurable intelligent surfaces channels are considered  \cite{GongHow2023}.  
The channel-oriented REK incorporates environmental information, departing from traditional statistical or deterministic channel models.
\textbf{Feature-oriented REK:} 
This category of knowledge is intended to describe the interactions between PEI from various feature dimensions and their impact on the channel. 
It includes standard data feature extraction, structured features based on graph relationships or other relational representations, semantic features that mine feature meanings, etc.
\textbf{Task-oriented REK:} 
This category of knowledge aims to describe which interactions among PEI directly influence communication tasks. 
Making direct decisions for communication tasks from environmental data is challenging because their relationship is indirect despite their direct connection to channel parameters.
More advanced knowledge representation methods are still being explored to achieve this.

\subsection{Dual interaction mechanism}
\begin{figure}[!t]
\centerline{\includegraphics[scale=0.35]{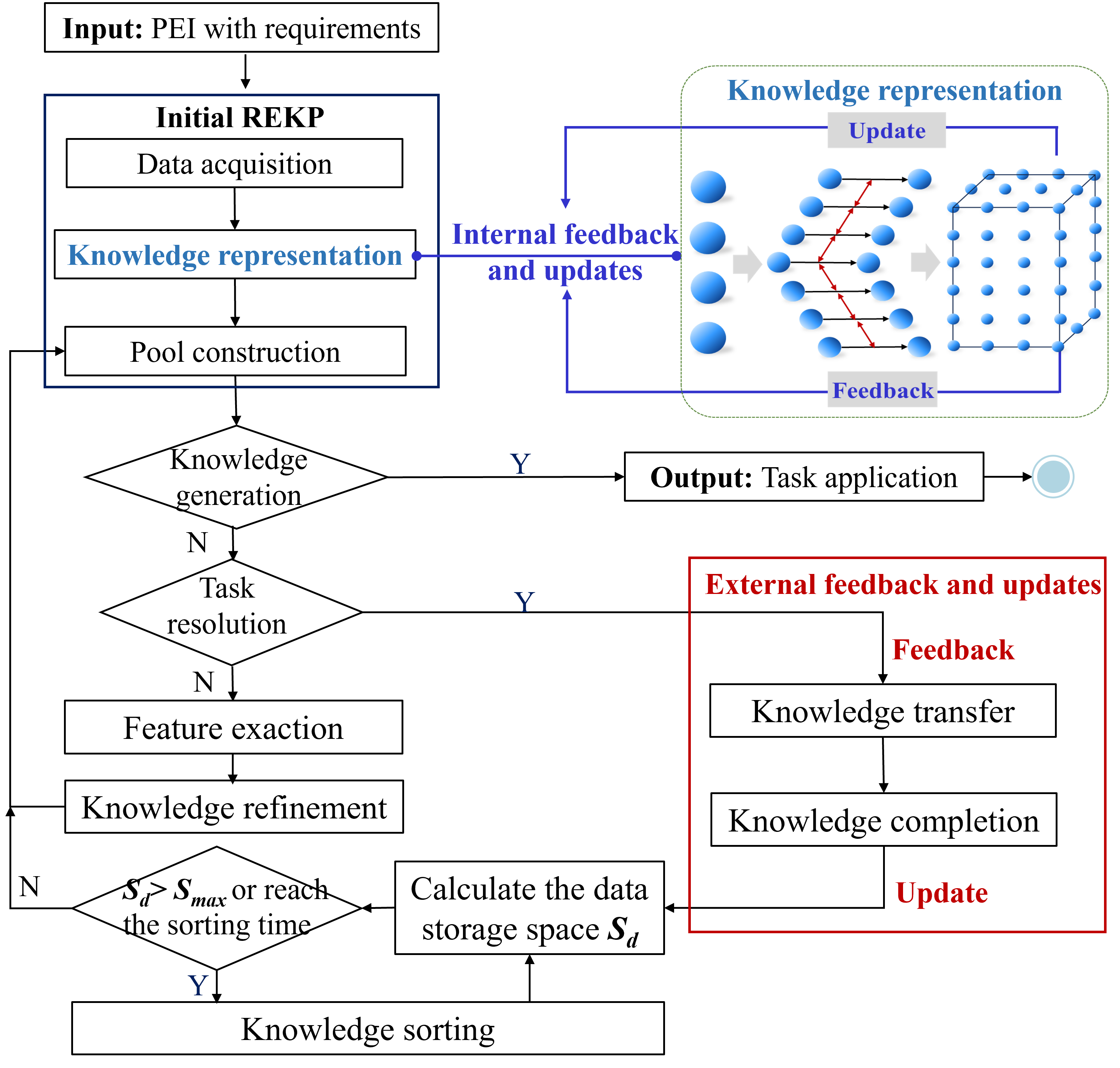}}
\caption{Process description of knowledge updating based on the dual interaction mechanism.}
\label{Fig3}
\end{figure}

The REKP framework encompasses knowledge generation, refinement, transfer, completion, and sorting, facilitated by a dual interaction mechanism, as depicted in Fig. \ref{Fig3}.

After substantial knowledge representation and initial REKP creation detailed in Section \ref{secIIIB}, inputs comprising PEI and specific requirements are received, typically related to desired channel parameters or directives for base stations or UTs.
Using the initial REKP, the system assesses whether new knowledge can be generated.  
If not, two scenarios may unfold: 
(1) Existing knowledge already exists but not captured, prompting task resolution involving feature extraction and refinement.  
(2) Alternatively, entirely new knowledge introduced by the inputs may require external feedback.  
A shared underlying representation exists between traditional channel parameters or tasks and new features and new tasks \cite{ZhuIntelligent2023}. 
Knowledge is transferred from the initial REKP to accommodate the new inputs using transfer learning principles.
The updated knowledge undergoes a knowledge completion process, facilitating REKP updates.
When storage permits or the knowledge sorting threshold is not met, updated knowledge is directly added to the pool.  
Otherwise, knowledge sorting is conducted according to similarity and utilization frequency criteria to remove outdated information.

From an overall perspective, to successfully employ REKP, it is essential to ensure the synergistic operation of knowledge scalability, communication tasks, and resources.
Resources encompass Computational capacity as well as communication resources such as spectrum, bandwidth, and time slots. 
Three components adaptively adjust based on PEI and REK to accomplish communication tasks like beam prediction, beam selection, and blockage prediction using minimal communication resources.
Meanwhile, hardware support is also required.
Hardware encompasses an array of equipment for acquiring PEI and channel data and environment reconstruction platforms \cite{MiaoDemo2023}.  
High-performance computing resources require substantial RAM, CPUs with more than two cores, and GPU accelerators.  
An extensive data processing capability is needed, comprising high-speed storage, robust network bandwidth, and parallel computing frameworks.

\section{Numerical results of knowledge representation} \label{secV}
In this section, a path loss knowledge construction method of composite environment feature relationship is proposed.

\begin{figure}[!t]
\centerline{\includegraphics[scale=0.475]{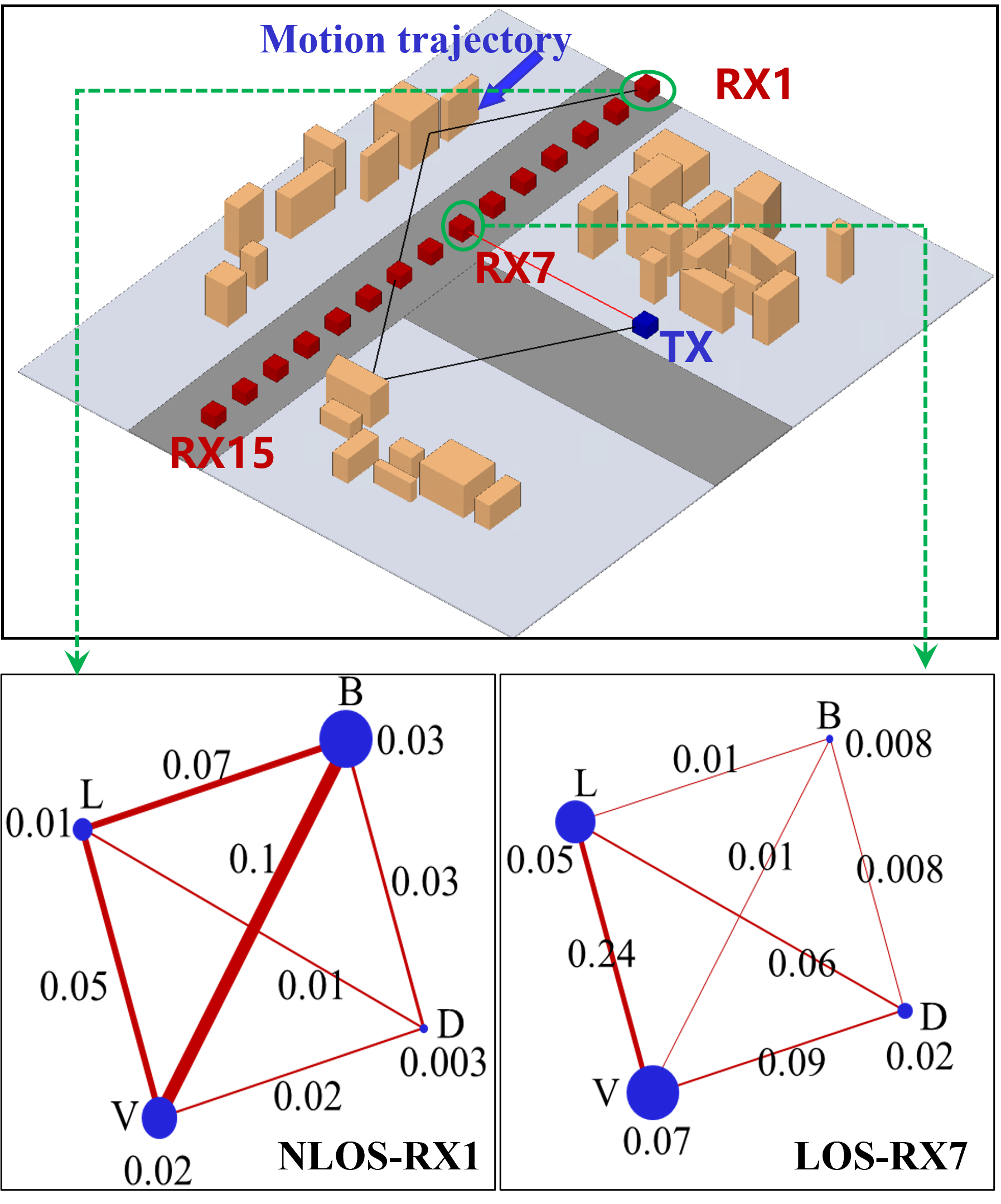}}
\caption{Simulation scenario of path loss knowledge representation and knowledge updates of RX1 and RX7.}
\label{Fig4}
\end{figure}

\begin{figure*}[!t]
\centerline{\includegraphics[scale=0.53]{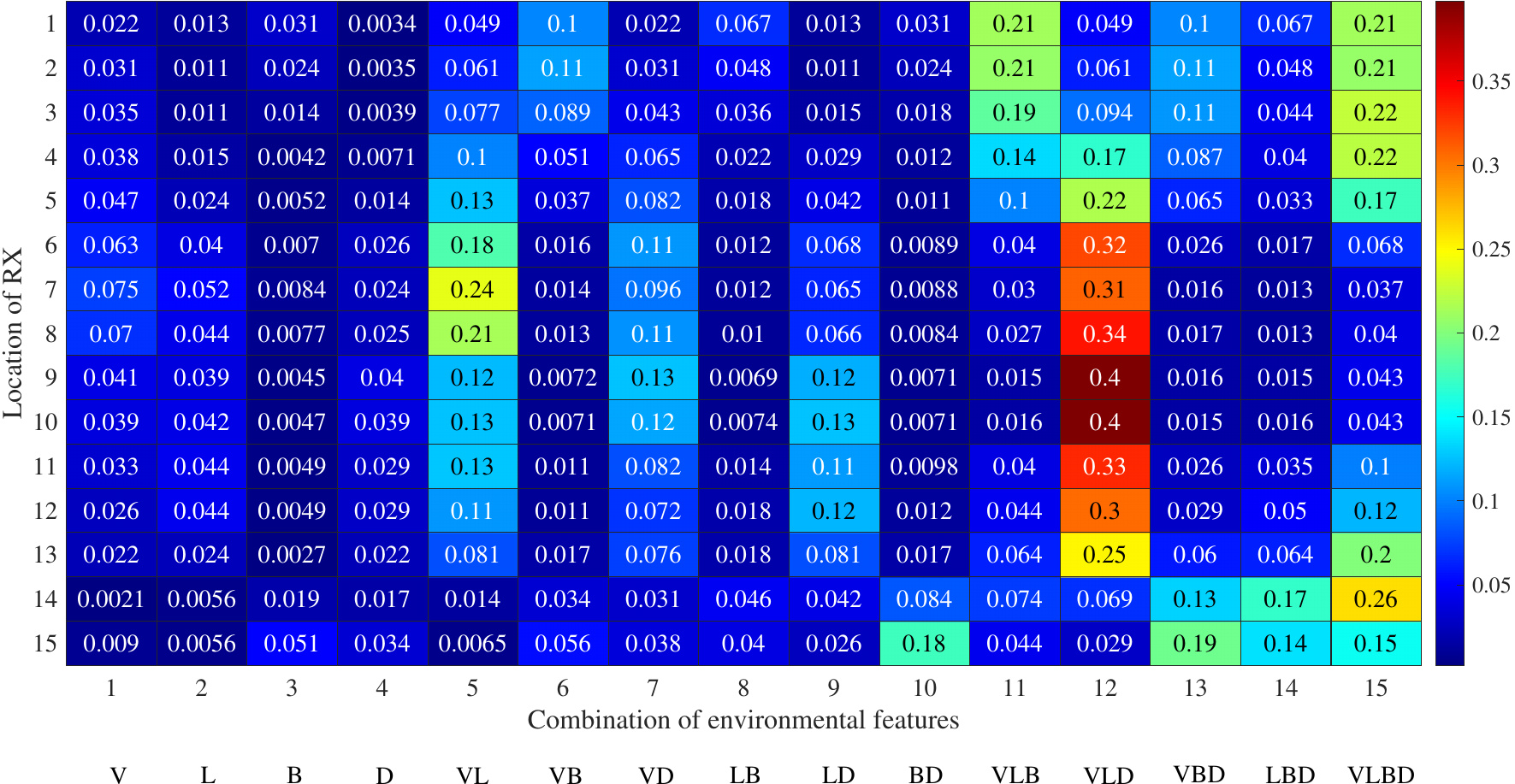}}
\caption{Quantified knowledge spectrum between the path loss and corresponding combined environment features.}
\label{Fig5}
\end{figure*}

The millimeter-wave communication system in the outdoor street environment is considered, as shown in Fig. \ref{Fig4}.
The simulation environment is generated using the commercial ray tracing tool Wireless Insite. 
Black and red lines indicate the most robust paths at the RX1 and RX7, respectively.
The TX is fixed at the blue square, and the RX follows a motion trajectory along the blue arrow, represented by red squares, covering 15 positions.
Vertex coordinates of scatterers and channel path loss are obtained.
Through vertex coordinates, PEI can be obtained, including three-dimensional coordinates of scatterers, TX, and RX, length, width, height of scatterers, occlusion for effective scatterers under NLOS, and distances from TX or RX to effective scatterers.
Based on the PEI, four environment features are extracted, namely, \textbf{location features $L$} related to three-dimensional coordinates of scatterers, TX, and RX, \textbf{volume features $V$} related to length, width, and height of scatterers, \textbf{blockage features $B$} and \textbf{distance features $D$}.
The impact weights of four environment features on path loss at the current location are individually learned utilizing the random forest. 
Based on these four environment features, there are 15 non-repetitive combinations, as illustrated in Fig. \ref{Fig5} on the x-axis coordinate. 
A graph representation of single and double feature relationships is constructed using the impact weights, the feature relationships graph for RX1 and RX7, as shown in the lower part of Fig. \ref{Fig4}.
Nodes and edges impact single-feature and double-feature relationships on path loss. 
Numbers represent the impact values of environment features on the path loss at the current location.
The higher the impact weights, the more significant the effect of the corresponding features on the fluctuation of path loss in the present environment.
It is found that during the movement from RX1 to RX7, the feature relationships are updated.

In Fig. \ref{Fig5}, the path loss knowledge is constructed, and the knowledge for all positions is quantified.
It can be observed that during the movement from RX1 to RX15, the patterns of path loss with single and multiple environment features under the same scenario are quantified, and the relationships between features are updated.
Knowledge updating is slow when RX remains in NLOS or LOS positions (as indicated on the y-axis from 1 to 4 or on the x-axis from 6 to 12).
At the NLOS positions, the numerical values of \emph{VLBD} are the largest, which indicates that four environmental features influence path loss at this location.
In densely scattered NLOS positions (e.g., RX1 and RX2 positions affected only by the largest-area cluster of scatterers), the impact of \emph{D} is minimal because the knowledge of \emph{VLB} and \emph{VLBD} is almost the same.
In LOS positions, the primary effect is from the three-feature relationship \emph{VLD}.
In reference \cite{SunEnvironment2022}, cumulative distribution functions (CDFs) of prediction errors produced utilizing the REKP methodology, convolutional neural networks (CNN), and artificial neural networks (ANN) are presented at 28 GHz.
Results show that the upper error of REKP is lower than CNN and ANN.
At 80\% CDF point, our idea provides a significant reduction compared with the other methods with more than 1 dB at 28 GHz.
Compared to previous work, the proposed knowledge construction method identifies which environment features impact path loss in dynamic scenarios and provides specific numerical values of the relationships between environment features.

\section{Open issues for DTC achievement}\label{secVi}
There are still open issues to achieving high performance, low cost, and reliability in 6G WCS.
REKP serves as an effective enabler in the following capabilities for achieving DTC.

\textbf{High performance:}
New developments and technologies in 6G build upon the foundations of 5G.  
For instance, the channel models used in the industrial Internet of Things (IIoT) still rely on traditional statistical theories from 5G.  
The controllability of REKP ensures the accuracy of techniques enabled by DTC.  
By analyzing the feature relationship between PEI and channel data, tailored knowledge is acquired for specific scenarios.  
Through knowledge transfer, twin channels are generated to facilitate communication tasks.

\textbf{Low overhead and cost:}
Low overhead is constantly pursued in WCS design.
The complexity is exceptionally high in massive multiple input multiple output (MIMO) configurations with hundreds or even thousands of antennas at the base station, which challenges traditional pilot-based channel feedback. 
The interpretability of REKP provides regular patterns between PEI and electromagnetic variations.
By sensing the behavior of UTs in the PW, the channel state is extrapolated, and the pilot or channel state information feedback overhead is reduced.

\textbf{Trustworthy DTC:} 
A digital twin must enable security. It must be as open, and transparent as possible, and built using legitimately good-quality data.
DTC implementation requires a large amount of data based. 
Currently, the quality of the associated data of environment and channel needs to be completed, irregular, or unpublicized.
Building an open-source, standardized radio environment database for DTC is one of the directions that needs to be explored.
Furthermore, using knowledge representation and the dual interaction mechanism in REKP to achieve data deduction and provide endogenous data is also a method of legitimate data generation.

\section{Conclusion} \label{secVI}

In this article, the REKP framework is proposed, which is a general framework aimed at in-depth knowledge representation and storage between environment information and channel data.
Firstly, the current state of the digital twins for the channels is summarized, and the results are compared to the key indicators of the digital twins.
Secondly, the construction method of REKP is introduced, including data acquisition, knowledge representation, and pool construction, explaining the dual interaction mechanism in the process of REKP updating.
Then, a path loss knowledge construction method of composite environment feature relationship is proposed, and numerical results for dynamic environment path loss prediction are based on REKP.
Finally, some open issues for DTC achievement are discussed from the perspective of high performance, low overhead and cost, and trustworthy DTC.
In future work, we will continue to explore diverse radio environment knowledge representation methods, which include intrinsic inference from channel data, environment, and communication semantic deconstruction, etc.

\section{Acknowledgment}
This work is supported by the National Key R\&D Program of China (No. 2023YFB2904803), the National Natural Science Foundation of China (No.92167202), the National Science Fund for Distinguished Young Scholars (No.61925102), and BUPT-CMCC Joint Innovation Center.

\begin{IEEEbiographynophoto}{JIALIN WANG (wangjialinbupt@bupt.edu.cn)}
received the B.S. degree from the Dalian University of Foreign Languages in 2018, and the M.S. degree from the North China University of Science and Technology in 2021. She is currently pursuing the Ph.D. degree in the Beijing University of Posts and Telecommunications (BUPT). Her current research interests include wireless channel measurement and modeling, channel prediction, machine learning.
\end{IEEEbiographynophoto}

\begin{IEEEbiographynophoto}{JIANHUA ZHANG (jhzhang@bupt.edu.cn) }
received the B.S. degree from the North China University of Technology in 1994 and the Ph.D. degree from the BUPT in 2003, where she is currently a professor. Now she is the Chairwomen of China IMT-2030 tech group - channel measurement and modeling subgroup and works on the 6G channel model. Her current research interests include Beyond 5G and 6G, artificial intelligence, and data mining, especially in mmWave, THz, and massive MIMO channel modeling.
\end{IEEEbiographynophoto}

\begin{IEEEbiographynophoto}{YUXIANG ZHANG (zhangyx@bupt.edu.cn) }
received the B.S. degree from the Dalian University of Technology in 2014 and the Ph.D. degree from the BUPT in 2020. From 2018 to 2019, he was a Visiting Scholar with University of Waterloo. He is now a Post-doctoral researcher. His current research interests include transmission techniques for physical layer, OTA testing and etc.
\end{IEEEbiographynophoto}

\begin{IEEEbiographynophoto}{YUTONG SUN (sun\_yutong@bupt.edu.cn) }
received the B.S. degree and M.S. degree from Changchun University of Science and Technology (CUST) in 2015 and 2018. Since 2020, she has been pursuing the Ph.D. degree in BUPT. Her current research interests include channel prediction, computer vision,  machine learning and etc. 
\end{IEEEbiographynophoto}

\begin{IEEEbiographynophoto}{GAOFENG NIE (niegaofeng@bupt.edu.cn) }
received the B.S. degree in communications engineering and the Ph.D. degree in telecommunications and information system from the BUPT, in 2010 and 2016, respectively. His research interests include SDN over wireless networks and key technologies in 5G/B5G wireless networks.
\end{IEEEbiographynophoto}

\begin{IEEEbiographynophoto}{LIANZHENG SHI (shilianzheng@bupt.edu.cn) }
received the B.S. degree from Yanshan University in 2023. He is currently pursuing the M.S. degree in information and communication engineering the BUPT. His current research interests include channel modeling, channel prediction, deep learning and etc.
\end{IEEEbiographynophoto}

\begin{IEEEbiographynophoto}{PING ZHANG (pzhang@bupt.edu.cn) }
received his M.S. degree from Northwestern Polytechnical University, Xi’an, China, in 1986, and his Ph.D. degree from BUPT, Beijing, China, in 1990. He is a Professor at BUPT, and the Director of the State Key Laboratory of Networking and Switching Technology. He is also an Academician with the Chinese Academy of Engineering. His research interests mainly focus on wireless communications.
\end{IEEEbiographynophoto}

\begin{IEEEbiographynophoto}{GUANGYI LIU (liuguangyi@chinamobile.com) }
received the Ph.D. degree from the BUPT. He is currently Fellow and 6G lead specialist, China Mobile Research Institute, where he is in charge of the wireless technology research and development, including 5G and 6G.
\end{IEEEbiographynophoto}


\end{document}